# Magneto-Transport Properties of Exfoliated Graphene on GaAs

Mirosław Woszczyna, Miriam Friedemann, Klaus Pierz,
Thomas Weimann, and Franz J. Ahlers*

*Physikalisch-Technische Bundesanstalt, PTB*
*Bundesallee 100, D-38116 Braunschweig, Germany*

**Abstract**

We studied the magneto-transport properties of graphene prepared by exfoliation on a III-V semiconductor substrate. Tuneability of the carrier density of graphene was achieved by using a doped GaAs substrate as a back-gate. A GaAs/AlAs multilayer, designed to render the exfoliated graphene flakes visible, also provides the required back-gate insulation. Good tuneability of the graphene carrier density is obtained, and the typical Dirac resistance characteristic is observed despite the limited height of the multilayer barrier compared to the usual $SiO_2$ oxide barrier on doped silicon. In a magnetic field weak localization effects as well as the quantum Hall effect of a graphene monolayer are studied.

* Corresponding author: franz.ahlers@ptb.de

## I. Introduction

Graphene is probably the most exciting material discovered in the last few decades. Its unusual mechanical and electronic properties suggest numerous applications, most of them based on the electronic transport properties of the material. Being the first material, however, which consists only of surface and not of bulk, these properties are strongly influenced by the choice of the supporting substrate, by potential contamination layers on the substrate, as well as by adsorbates or capping layers placed on top of the graphene mono- or few-atomic layers. Electronic mobility, for example, one of the key parameters determining the transport properties, is highest for exfoliated graphene. Especially freely suspended graphene[1], or graphene placed on the host substrate boron nitride[2] which is well lattice matched, exhibit extremely high mobilities, allowing the observation of the fractional quantum Hall effect.[3] Bottom-up methods like the growth of graphene on SiC, or CVD growth on metal surfaces can provide much larger graphene sheet sizes, but usually with inferior mobility[4] up to now. Since CVD growth is usually performed on catalytically active metal surfaces, the transfer of the graphene layers to an insulating substrate is required to exploit their electrical properties. However, the substrate choice does not necessarily have a deteriorating effect only. It can also be exploited to obtain new functionality by taking advantage of a combination of another material's specific properties with those of graphene. Two examples are the above-mentioned combination of boron nitride and graphene, or recent theoretical suggestions to combine graphene with ferromagnetic material[5-8] in order to enable spin polarized transport in graphene via the ferromagnetic proximity effect. For the latter, the III-V semiconductor GaAs:Mn could be a suitable choice. Also the study of graphene by more exotic probing methods, like e.g. surface acoustic waves, becomes possible when the material is placed on a piezoelectric material like GaAs.

Up to now, however, relatively few studies have been reported on graphene on III-V semiconductor surfaces. This is mainly due to the lack of a simple device fabrication method on these materials, as is available for Si/$SiO_2$ substrates.[9] Only a few studies have focussed on graphene on GaAs.[10-12] None of them is on electrical transport, despite some advantages which this electronic material, which is the best known next to Silicon, could have. The superior surface quality of GaAs compared to the usually very rough thermally oxidized $SiO_2$ should positively influence the transport properties of graphene, and its higher dielectric constant would cause shorter screening lengths of charged impurities. Understanding the properties of graphene on GaAs could be a first step towards combining it with Mn-doped GaAs, eventually aiming at spintronic applications. In this paper we present , after our earlier demonstration[12] that graphene can be exfoliated and made as easily visible on GaAs as on $SiO_2$, the first study of the electronic, and especially magneto-transport properties of this material combination. The tuneability of the carrier density is achieved by using a doped GaAs substrate as a back-gate, whereby the GaAs/AlAs multilayer needed to achieve optical visibility also provides the insulating barrier between the back-gate and graphene.

## II. Device Preparation

Specially layered GaAs/AlAs substrates for graphene exfoliation and identification were grown by molecular beam epitaxy, as described in [12]. While the layer sequence shown in Fig. 1(a) provides good visibility in the green spectral region, the energy band structure of the layers, see Fig. 1(b), indicates that it should also function as an insulating barrier. Its electrical breakdown voltage is expected to be lower than for an $SiO_2$ insulator of similar thickness, but the higher dielectric constant of GaAs/AlAs partly relaxes this constraint since the gate voltage for a given graphene carrier density change can be three times smaller.

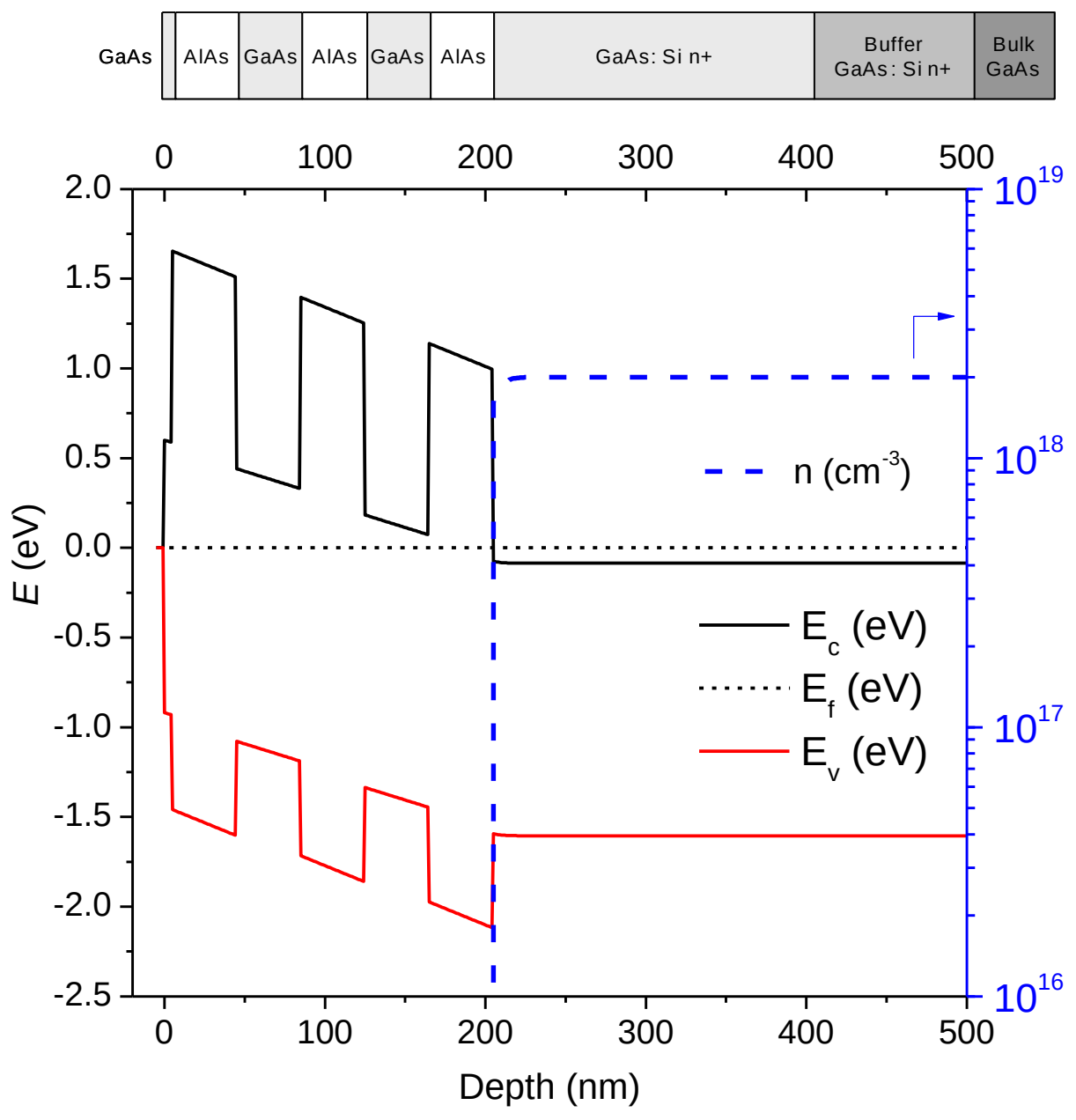

Fig. 1. (Colour online) (a) A schematic diagram of the fabricated substrate layer sequence. (b) A band structure simulation of the GaAs/AlAs substrate. The upper trace indicates conduction band energy, the lower one shows valence band energy, the dotted line shows the Fermi energy level. The dashed line shows the theoretical carrier density in the heterostructure.

The detailed growth sequence for the substrate was as follows: In a standard III-V solid source molecular beam epitaxy (MBE) system 300 nm of Si-doped $n^+$-GaAs were grown as a starting layer to extend the $n^+$-type GaAs(100) back-gate substrate. This was followed by 20 nm of undoped GaAs, to suppress Si segregation into the upper layers. During this growth step the substrate temperature was increased from 580°C to 640°C in order to obtain smooth AlAs/GaAs interfaces. Next, the dielectric mirror, important for graphene visibility, was grown, consisting of three AlAs layers and two GaAs layers, each of 40 nm thickness. To prevent the oxidation of the uppermost AlAs layer, the structure was capped by 5 nm of GaAs. The roughness of the fabricated heterostructures was checked with atomic force microscopy, and over several scanned areas, a mean *RMS* roughness of about 0.3 nm was determined, a factor of two lower than for the thermally oxidized $SiO_2$ substrates, which we had studied in parallel.

Graphene was prepared on this substrate by the exfoliation method, in practically the same fashion as usually employed on a Si/$SiO_2$ substrate.[9,13] The resulting flakes were located in an optical microscope using green light for contrast enhancement.[12] The method is sensitive enough to distinguish graphene mono- and bi-layers. In a next step devices were fabricated using electron beam lithography and PMMA masks. The geometrical definition of Hall bars was achieved by dry etching of graphene in an argon/oxygen plasma. In a final step, metal contacts (100 nm Au on top of 10 nm Ti) were made by evaporation and lift-off. Typical Hall bar dimensions were 20 µm by 10 µm, but occasionally much bigger flakes were obtained with this technique (Fig. 2).

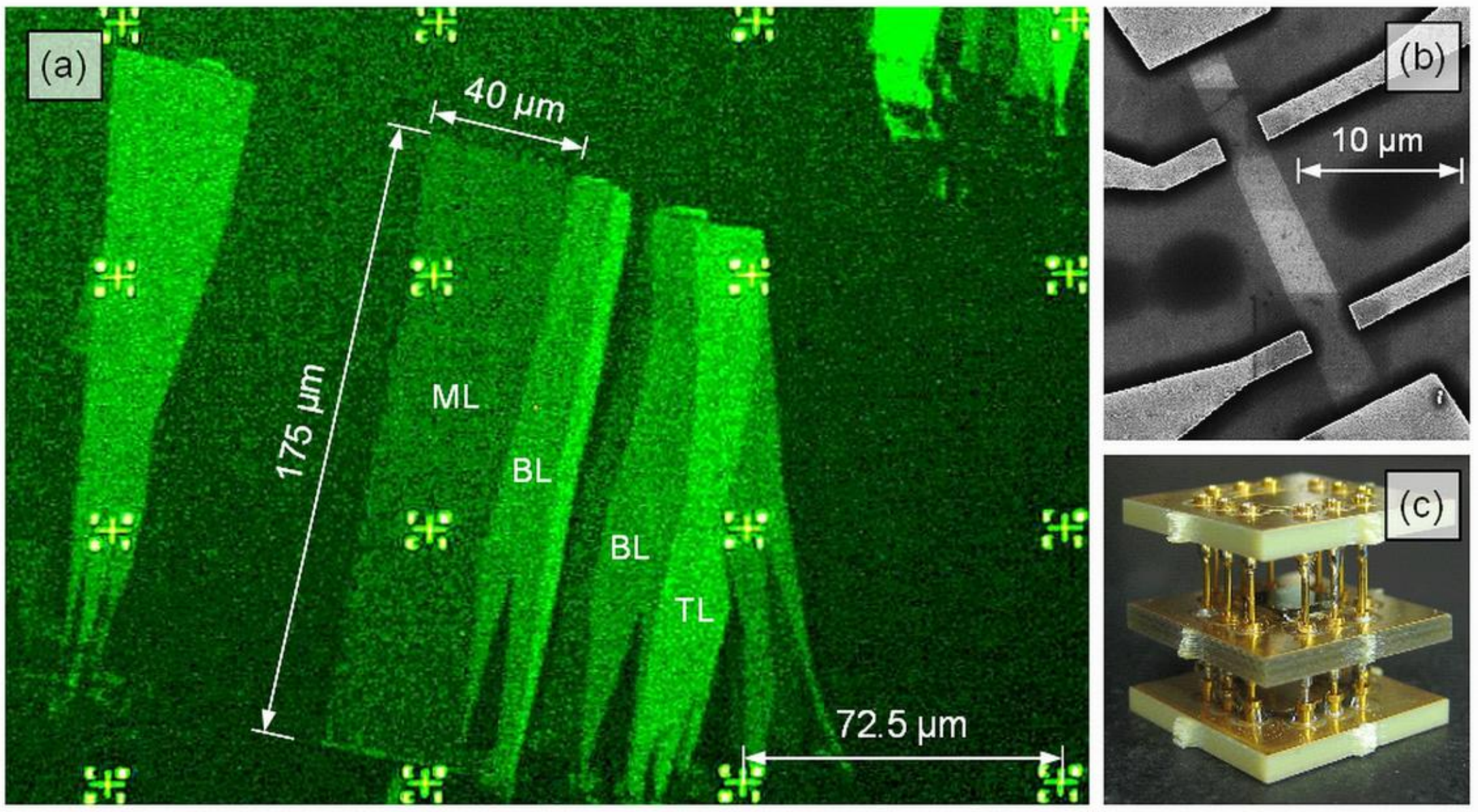

Fig. 2. (Colour online) (a) Graphene flakes deposited on a GaAs substrate. Mono- (ML), bi- (BL) and tri-layers (TL) of graphene can be easily distinguished by an optical microscope when equipped with a 550 nm green filter for contrast enhancement [12]. (b) Example of scanning electron micrograph of a graphene Hall bar on a GaAs substrate. (c) A picture of an assembled device in protective housing. The graphene device (the middle part) is mounted in a TO8 compatible chip carrier which is sandwiched by two PCB carriers with all contacts short. This approach keeps all graphene terminals at the same electrical potential and prevents damaging by electrostatic discharge, for instance, during wire bonding or mounting on to a measurement setup.

## III. Electrical Transport Properties

As already mentioned the accessible range of back-gate voltages is limited by current leakage due to electrical breakdown between the GaAs substrate and graphene. The *IV*-curves in Fig. 3 illustrate typical leakage in our devices. At liquid helium temperature leakage starts at a forward bias of –1 volt (back-gate negative with respect to graphene), while under reverse bias up to +5 volts can be applied, giving a total range of 6 volts for carrier density tuning.  At room temperature the leakage is larger, but it still allows a rough check of device integrity.

Before cooling down, the samples were kept in vacuum (approx. $10^{-6}$ hPa) at room temperature for two hours. At base temperature (15 mK) and using dc current levels of 100 nA, two-terminal longitudinal resistance measurements were performed (Fig. 4a). The conductivity was determined upon varying the graphene carrier density via the back-gate voltage and the result is shown in Fig. 4b. Very favourably we found that - already in the as-prepared state - the charge neutrality point (CNP) of graphene was always at back-gate voltages very close to zero, in stark contrast to exfoliated graphene on $SiO_2$, where - without special treatment - voltages of typically around +40 V or more are required to tune the carrier density to zero. Therefore, the limited back-gate voltage of only +5 volts presented no hindrance for the use of graphene on GaAs. We studied each device in several cool-down cycles and found that the CNP was slightly shifted between cool-downs. The data below are from one device, but were collected in different cool-down runs.

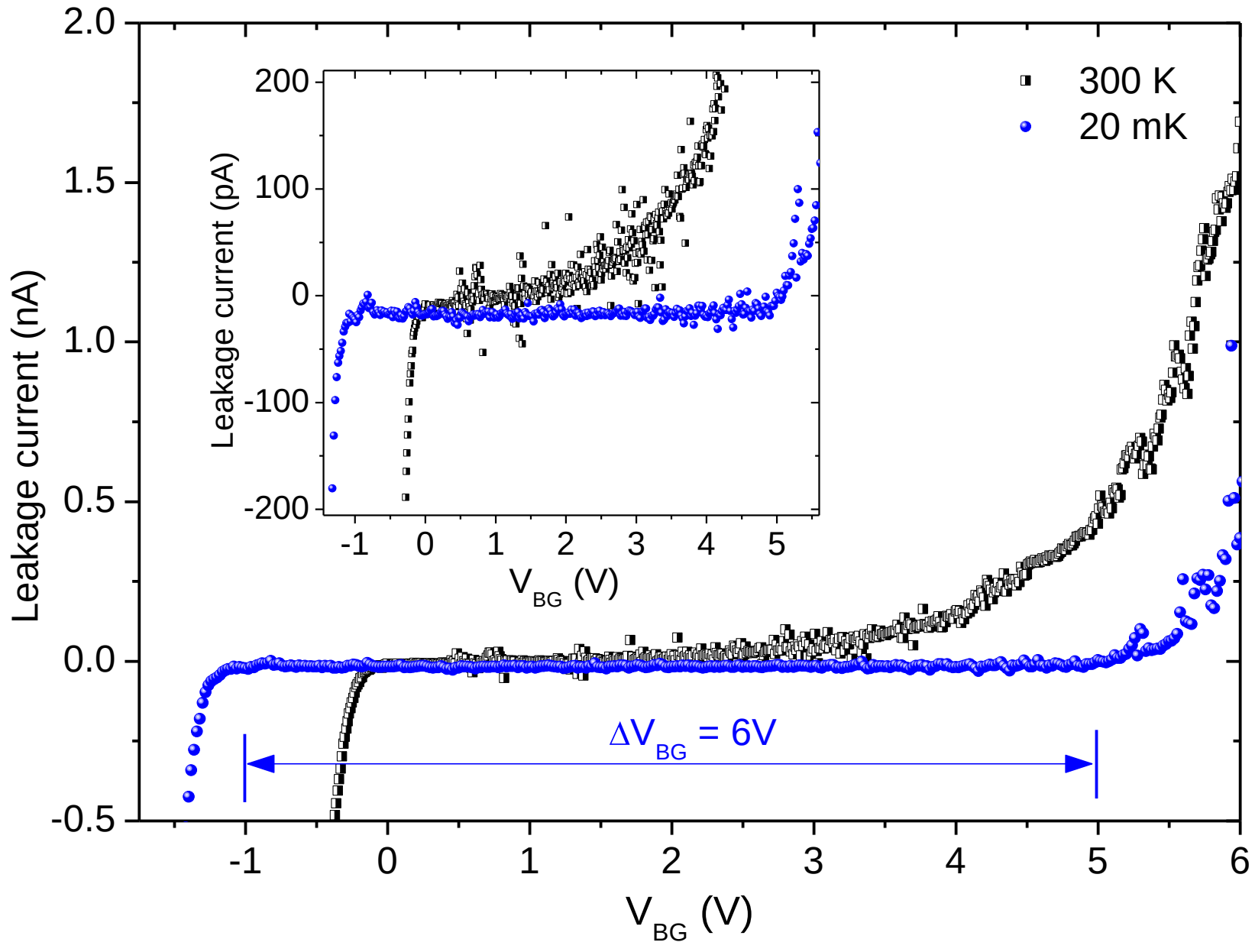

Fig. 3. (Colour online) Leakage current between graphene and a back-gate of a GaAs/AlAs heterostructure. The usable range of back-gate voltages is significantly increased at low temperatures. The square black points represent leakage current at room temperature, and the circular blue points that at low temperature ($T$ = 20 mK), respectively.

At different back-gate voltages the carrier density $n$ was determined from Hall measurements. A linear dependence $n = \alpha(V_{BG} - V_0)$ was observed (Fig. 4c), with $\alpha$ = $28.9\times10^{10}$ $cm^{-2}V^{-1}$, corresponding to a gate capacitance per unit area $C_G = \alpha e$ ($e$ is the elementary charge) of 0.46 $nF/mm^2$. The values are in good agreement with a theoretical estimate based on a plate capacitor model, which predicts $\alpha = 30\times10^{10}$ $cm^{-2}V^{-1}$ and $C_G$ = 0.48 $nF/mm^2$. In the calculation an effective static relative permittivity of $\kappa_{eff}$ = 11.2 was used for the multilayer substrate, compared to the value $\kappa_{SiO_2}$ = 3.9

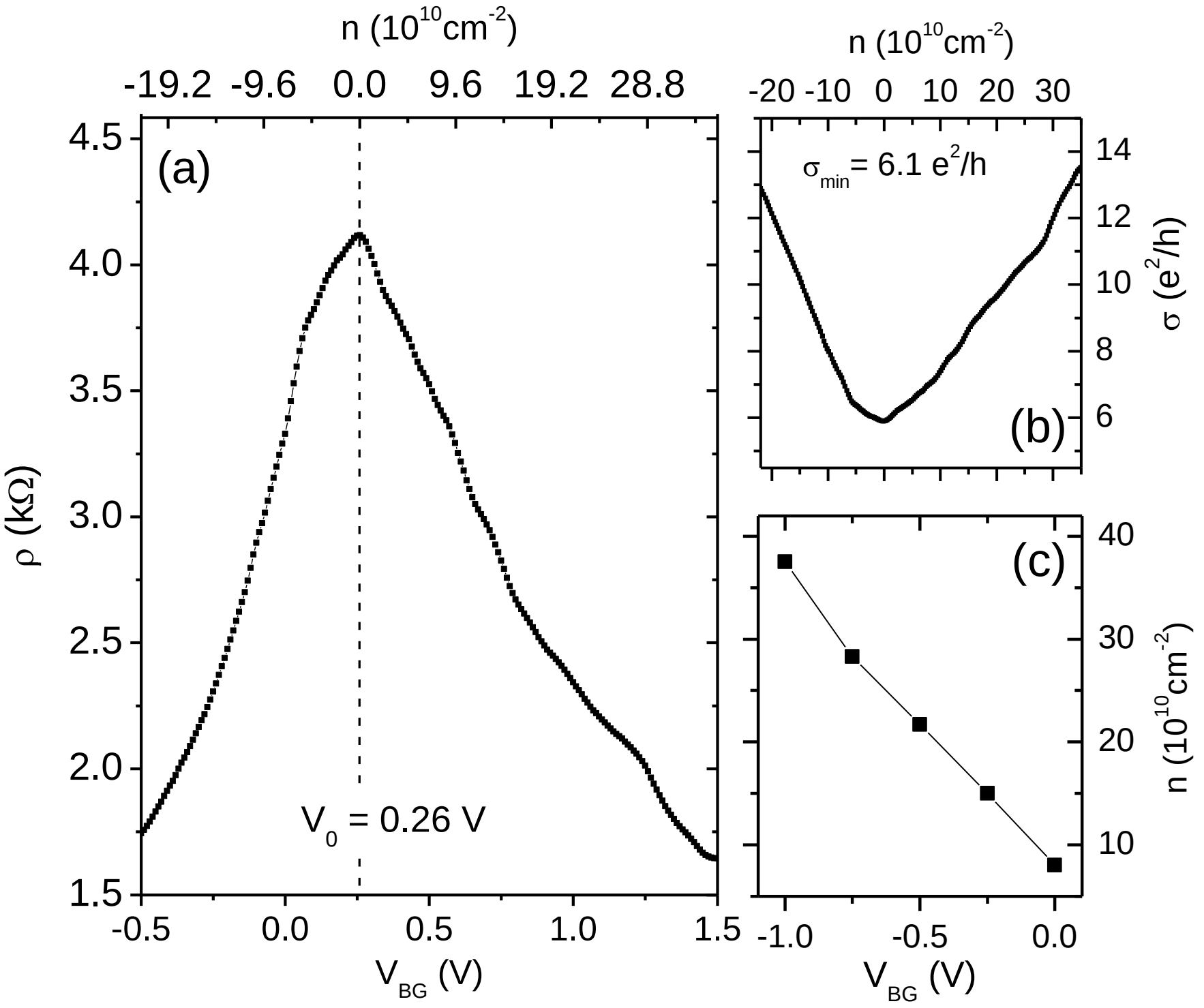

Fig. 4. (a) Graphene device resistivity change as a function of back-gate voltage. The charge neutrality point (CNP) for this sample occurred at $V_0$=0.26 V at $T$ = 15 mK. The upper horizontal scale gives the carrier concentration determined according to the description under (c). (b) Conductivity in dependence on graphene carrier concentration. (c) Carrier concentration determined from Hall measurements. A linear relation is observed with proportionality constant of $28.9\times10^{10}$ $cm^{-2}V^{-1}$. (For 300 nm $SiO_2$ dielectric layer, the respective value is $7.2\times10^{10}$ $cm^{-2}V^{-2}$ [13]).

[14] for $SiO_2$. Correspondingly a $SiO_2$ layer of 300 nm thickness exhibits values of $\alpha_{SiO_2}$ = 7.2×10$^{10}$ cm$^{-2}$V$^{-2}$ and $C_{GSi0_2}$ = 0.11 nF/mm$^2$.[15] For the minimum conductivity at the CNP we obtained $\sigma_0$ ≈ 6.1 $e^2/h$, and the field effect mobility, defined by $\mu = C_G^{-1}\, d\sigma/dV_{BG}$, was about 1.2 T$^{-1}$ for holes and 1.0 T$^{-1}$ for electrons.[16] Contact resistance was approximately 1 kΩ at low temperature for a typical contact size of 5×3 µm². These values compare favourably to the typical mobility of 0.4 … 0.7 T$^{-1}$, which were obtained with practical identical processing on our Si/$SiO_2$ substrates.

In order to estimate the charged impurity density in the GaAs substrate and the residual carrier density carrying the current at the CNP, we applied the self-consistent theory for graphene transport described in [17]. From this we obtained, for the impurity carrier density $n_{imp} = 2e/h\mu F_l(r_s)$, a value of approximately 8.6×10$^{11}$ cm$^{-2}$, and for the residual carrier density $n_* = \sigma_{min} h F_l(r_s) n_{imp}/2e^2$ a value of 1.3×10$^{11}$ cm$^{-2}$. In these expressions $r_s = 2e^2/\hbar v_F(\kappa_1+\kappa_2)$ = 0.33 is an interaction parameter which depends on the dielectric constants of the material sandwiching the graphene flake, $v_F$ = 1.1× 10$^{10}$ m/s is the Fermi velocity, and
$F_l(r_s) = \pi r_s^2 + 24 r_s^3(1-\pi r_s) + 16 r_s^3(6r_s^2-1)\arccos(1/2r_s)/\sqrt{(4r_s^2-1)}$ is a long-range scattering parameter.[18] In our case $\kappa_1$ = 11.2 for GaAs substrate and $\kappa_2$ = 1 for vacuum. From the CNP shift $V_0$ of 0.26 V, we obtain a value $n_0 = C_G V_0/e$ = 7.8×10$^{10}$ cm$^{-2}$, whereas from the model, a theoretical CNP shift of $V_{0th} = \sqrt{\pi}\, \hbar v_F n_{imp}/2e\sqrt{n_*}$ of 0.15 V would be predicted, corresponding to a mean carrier density of 4.6×10$^{10}$ cm$^{-2}$. All these values are in reasonable agreement. It should be pointed out that a bigger dielectric constant of the substrate leads to a lower interaction parameter $\alpha$. According to the theory[17] this leads, at a given charged impurity concentration, to a higher device mobility, since $F_l(r_s)$ is a monotonically decreasing function. (For $SiO_2$ $r_s \approx 0.81$, and for the GaAs/AlAs substrate $r_s$ is approx. 0.33). The above calculations rely only on the theory of long-range scattering on charged impurities in graphene around the CNP. Recent experiments and theoretical work suggest that resonant scatterers also play an important role in transport, and limit graphene’s mobility.[19,20,21]

## IV. Magneto-transport Measurements

To fully characterize the electrical transport properties and the quality of the Hall quantization, measurements were performed in magnetic fields up to 13 T for both field polarities at $T$ = 50 mK. The longitudinal and Hall resistance measurements shown in Fig. 5 were taken at a hole carrier density of 5.3×10$^{11}$ cm$^{-2}$ (corresponding to -1 V back-gate voltage in this cool-down cycle) using an 8½ digit DVM and a 1 µA current source, a combination which was carefully calibrated against a 12.9 kΩ standard resistor. The field sweep rate was 0.5 T/min. The curves exhibit wide and clearly visible quantized Hall plateaus beyond ±6 T. They correspond to Landau level filling factor $\nu = 2$ and confirm the monolayer character of the flake. Quantized resistance features at filling factors of 6 and 10 were barely noticeable, however. The insets (c) and (d) of Fig. 5 present zoom-in views of the Hall resistance at the plateaus in comparison with the theoretical resistance value $h/2e^2$. We averaged the measured Hall resistance values over an interval $\Delta B = \pm 1\,\mathrm{T}$ at the approximate plateau centres located at -10 and +10 T, and calculated the relative deviations $\delta R_{\mathrm{H}}$ from the theoretical value. Relative deviations of +3.3 ppm and +4.0 ppm were obtained for the two field directions, both within the total measurement uncertainty which amounted to 7.7 ppm. In order to check whether this could already indicate a possible deviation from exact quantization caused by a coexistence of electrons and holes, we further studied the quantization quality at reduced carrier densities, i.e. close to the graphene charge neutrality point and in magnetic fields below $B$ = 6 T (Fig. 6). Achieving such a low field operating regime is an important design goal for graphene based QHE resistance standards,[22] because in that regime, one of the fundamental differences of graphene compared to GaAs based quantum Hall resistance standards can be exploited, namely the square-root dependence of the Landau-level splitting on a magnetic field. The relative advantage of graphene over the conventional systems increases the lower the magnetic field, and, hence, the carrier density are. While it has been shown recently that graphene grown on SiC and covered by organic top layers[23] can be tuned to rather low carrier densities and still exhibit good carrier density homogeneity, it is not clear whether the same can be achieved with exfoliated graphene on other substrates.

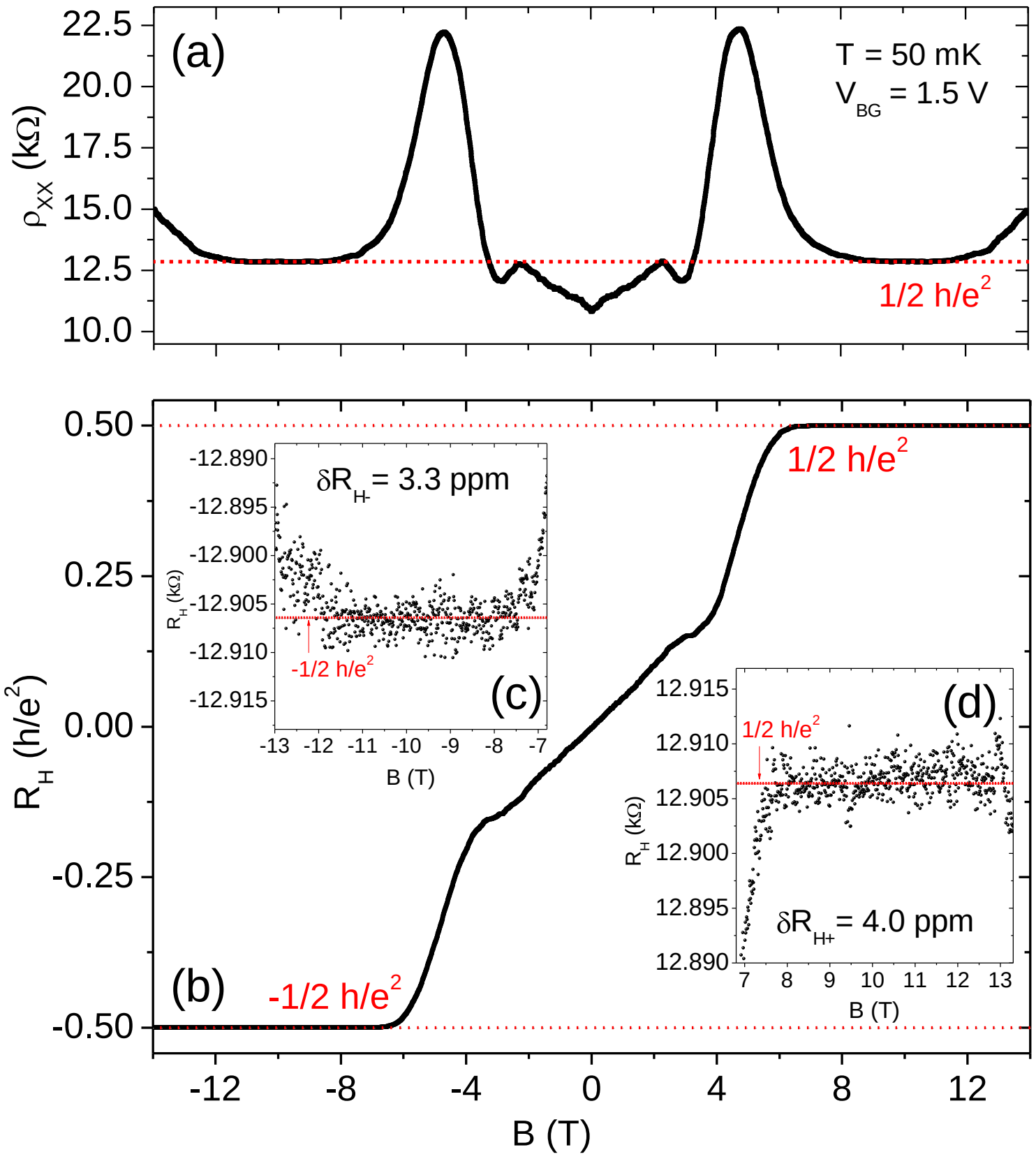

Fig. 5. (Colour online) Magneto-transport properties of graphene. (a) Two-terminal longitudinal resistivity and (b) Hall resistance for magnetic fields up to $B = 13$ T at temperature T = 50 mK and carrier concentration $n \approx -5.3\times10^{11}$ cm$^{-2}$. Dotted lines mark the value $h/2e^2$. Insets (c) and (d) show zoom-in views of the Hall resistance plateau. Mean resistance values were calculated from 200 data points from an interval of ±1 T around -10 T and +10 T, yielding a relative standard deviation of the mean of 7.7 ppm. Relative deviations from $h/2e^2$ of 3.3 ppm and 4.1 ppm at -10 T and +10 T, respectively, are within this uncertainty.

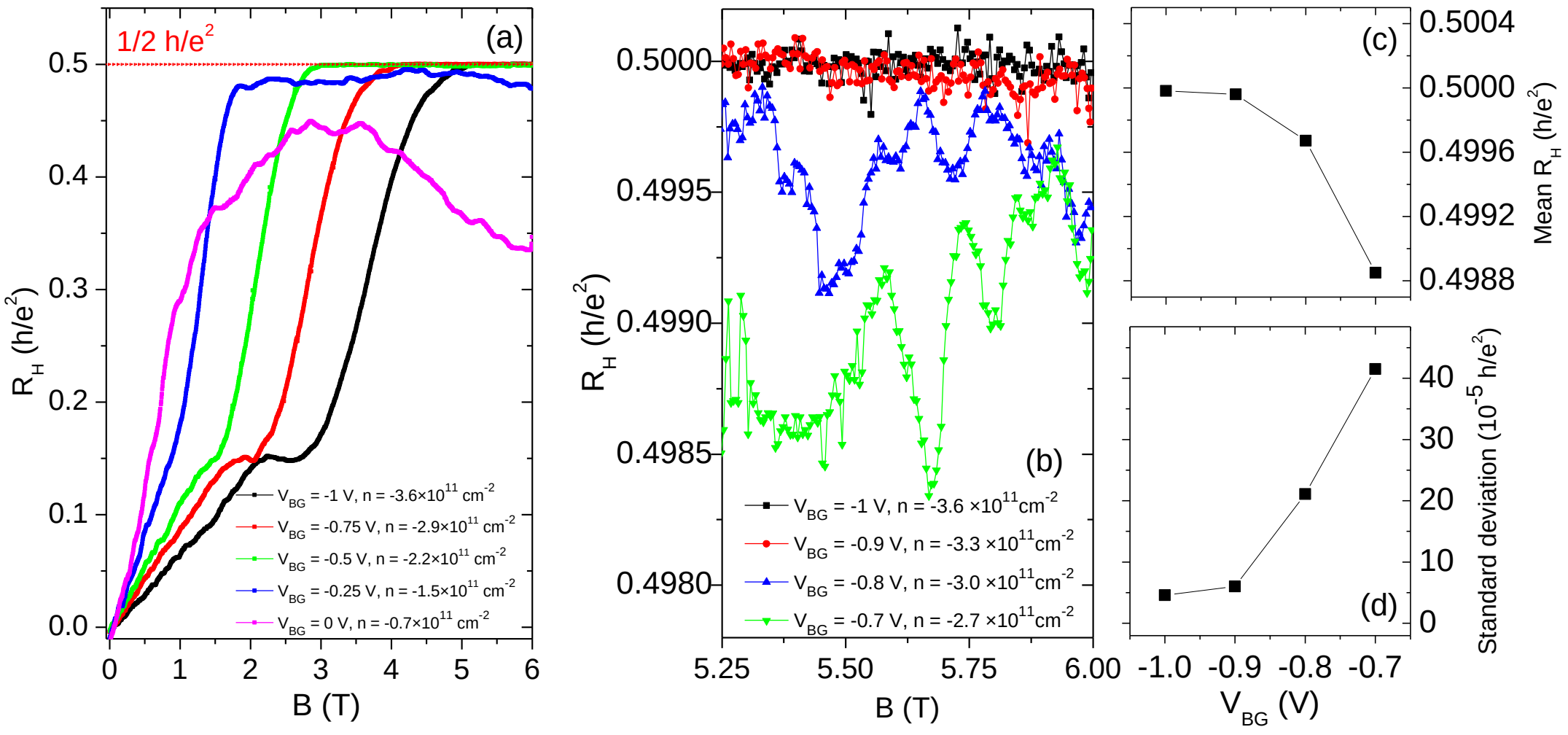

Fig. 6. (Colour online) (a) Quantum Hall resistances near the graphene charge neutrality point in moderate magnetic fields up to $B$ = 6 T at $T$ = 50 mK. A dotted line marks the value $h/2e^2$. (b) Hall resistance fluctuations for carrier concentrations varying from $-3.6\times10^{11}\,\mathrm{cm}^{-2}$ to $-2.7\times10^{11}\,\mathrm{cm}^{-2}$ in a magnetic field interval from 5.25 T to 6 T. Smaller carrier concentrations resulted in more pronounced distortions from a plateau. Simple measure of the distortion as a signal mean value and a standard deviation were utilized and shown in (c) and (d), respectively.

We performed this study in another cool-down cycle of the device where a back-gate voltage variation from -1 V and 0 V changed the hole carrier density from $3.6\times10^{11}\,\mathrm{cm}^{-2}$ to $0.7\times10^{11}\,\mathrm{cm}^{-2}$. The $\nu = 2$ Hall plateau is still observed for magnetic fields well below $B$ = 10 T, but at small carrier densities it becomes more and more distorted (Fig. 6a). The distortions are fully reproducible; such a behaviour was reported similarly for graphene on a $SiO_2$ substrate.[24,25] In Fig. 6b Hall resistance measurements in a magnetic field between 5.25 T and 6 T are presented for carrier densities in the range $3.6\times10^{11}\,\mathrm{cm}^{-2}$ to $2.7\times10^{11}\,\mathrm{cm}^{-2}$. Relatively small changes of $n$ resulted in strong, reproducible distortions, well above the noise level of the measure-ment in amplitude. The distortions are quantitatively described in Fig. 6c and d by plotting their mean value and standard deviation in dependence on the carrier density. For a hole carrier density of $3.6\times10^{11}$ cm$^{-2}$ the Hall resistance shows a relative deviation from the theoretical value by approximately 5 parts in $10^5$. This is already too high for an application as a quantum resistance standard. For even lower densities the deviation from exact quantization drastically increases, reflected by a monotonic decrease of the mean value and a significant increase of the standard deviation.

We speculate that the behaviour described reflects the electron- and hole-puddle formation - due to disorder[26] - in a graphene structure as proposed by Poumirol *et al.*[25] They observed similar phenomena in much higher magnetic fields of several tens of Teslas, and for higher carrier densities around $n = 10\times10^{11}$ cm$^{-2}$. It is widely accepted that the chemical potential variations within a graphene flake are mainly caused by non-homogeneously distributed charged impurities both under and above the flake.[27,28] Therefore, in the vicinity of the graphene charge neutrality point *n*-type and *p*-type carrier regions (puddles) form. In the case shown in Fig. 6 holes were the majority carriers, but both carrier types contributed to transport. The contribution of *n*-type charges increased more and more while reducing the total number of carriers by tuning the back-gate voltage towards the charge neutrality point. The minimum carrier density at which a still relatively precise Hall resistance quantization of 50 ppm is observed ($3.6\times10^{11}$ cm$^{-2}$ in our case), is in qualitative agreement with calculations based on a self-consistent theory for graphene transport.[27] Within this theory the residual carrier density $n_*$ of the device at the CNP is a statistical *rms* measure of carrier density fluctuations and it is considered to play the most important role in the graphene transport at the CNP.[27,29] Since the probability distribution function (PDF) of carrier density fluctuations is not a simple Gaussian[29], it is difficult to quantitatively assess the maximum range where electron-hole-puddle formation is significant. Ignoring this and approximating the PDF by a Gaussian (at a 99.7% confidence level), a maximum carrier density range of $n_{\mathrm{max}} = 3n_* = 3.9\times10^{11}$ cm$^{-2}$ can be estimated, close to the observed value of $3.6\times10^{11}$ cm$^{-2}$. Earlier investigations of the local electronic properties of a graphene flake with a scanning single electron-transistor[26] and a scanning tunnelling microscope[30,31] support this. It was reported[30,31] that carrier density fluctuations in graphene on $SiO_2$ are of the order of $(3.5 – 4.0)\times10^{11}$ cm$^{-2}$ at zero magnetic field. Those results were confirmed by graphene transport investigations in the quantum Hall regime.[32] Obviously graphene on GaAs behaves quite similarly to graphene on $SiO_2$, and our investigation is, to our knowledge, the first to report these observations.

## V. Weak Localization

In order to check for other similarities between graphene on GaAs and graphene on $SiO_2$, we also considered the weak localization effects of this material. In magnetic fields lower than 80 mT we studied the magneto-conductance of the graphene device at a temperature of 15 mK (Fig. 7). An applied magnetic field suppresses weak localization by altering the phase of electrons travelling in opposite directions along time-reversed paths, which leads to the observed change of the magneto-conductivity. We observed such a weak localization effect, superposed by reproducible variations reminiscent of universal conductance fluctuations. In addition, large and reproducible variations of the magneto-conductivity at different back-gate voltages $V_{BG}$ were observed (grey curves). To assess the mean weak localization contribution (black curve), we used the procedures employed in [33,34] and averaged several magneto-conductance traces (grey curves) taken at various $V_{BG}$ ranging from 0 to 1 volt in 0.1 volt steps. This voltage range corresponds to mean carrier densities between $8.3\times10^{11}$ cm$^{-2}$ and $6.3\times10^{11}$ cm$^{-2}$. The averaged magneto-conductivity curve can be described by a widely used weak localization model.[33] A fit of the model equation to our data is shown as a dashed trace in Fig. 7. However, conductivity corrections do not only depend on the dephasing time $\tau_\Phi$ due to inelastic scattering, but also on elastic scattering described by an intervalley scattering time $\tau_i$ and an intravalley scatter-ing time $\tau_*$. Intervalley scattering occurs on sharp defects and crystal structure imperfections, while intravalley scattering is mostly caused by dislocations with the size of the lattice spacing and graphene flake rippling.[35]

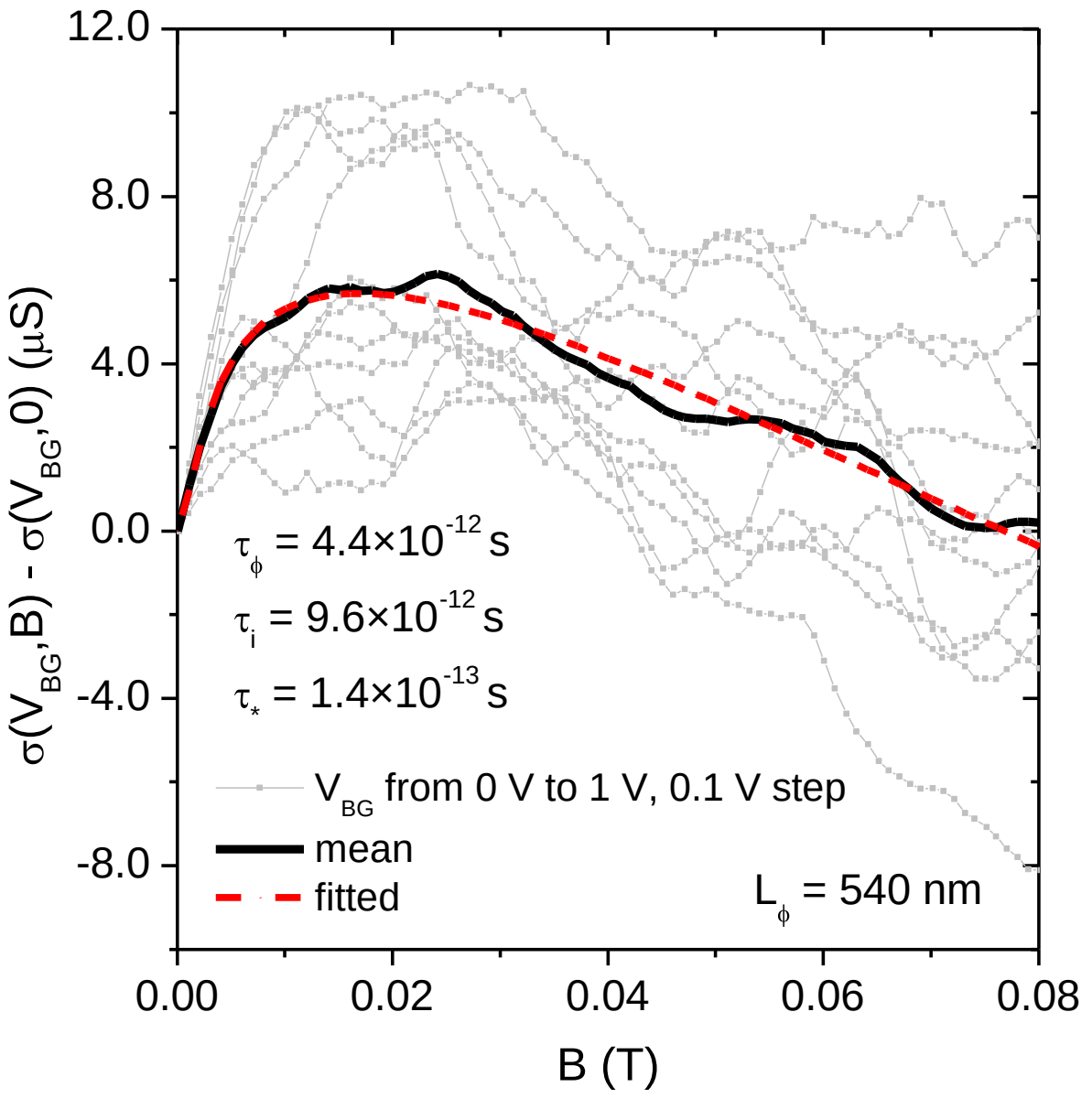

Fig. 7. (Colour online) Magnetoconductivity for various back gate voltages observed in the graphene device with magnetic fields up to 80mT (grey curves). The solid curve describes a mean magnetoconductivity for the $V_{BG}$ range from 0 V to 1 V with 0.1 V step, while the dashed one is a weak localization fitting curve. Charge neutrality voltage for this sample was 4.2 V and measurement temperature was 15 mK. Obtained dephasing time $\tau_\varphi$, intervalley scattering time $\tau_i$ and intravalley scattering time $\tau_*$ were approximately $4.4\times10^{-12}$ s, $9.6\times10^{-12}$ s and $1.4\times10^{-13}$ s. This corresponded to $L_\varphi \approx 540$ nm, $L_i \approx 800$ nm, and $L_* \approx 10$ nm, respectively.

The average conductivity correction $\Delta\sigma(B) = \sigma(V_{BG},B) - \sigma(V_{BG},0)$ due to the weak localization was evaluated utilizing the formula[34]

$$\Delta\sigma(B) = \frac{e^2}{\pi h}\left[F\left(\frac{B}{B_\phi}\right) - F\left(\frac{B}{B_\phi + 2B_i}\right) - 2F\left(\frac{B}{B_\phi + B_*}\right)\right]$$

where $F(x) = \ln(x) + \Psi(½+x^{-1})$, $B_{\varphi,i,*} = \hbar\tau_{\varphi,i,*}^{-1}/4De$ and $\Psi$ is a digamma function. Having determined the scattering times, one can estimate corresponding scattering lengths, according to $L_{\varphi,i,*} = (D\tau_{\varphi,i,*})^{1/2}$.[Fehler! Textmarke nicht definiert.,36] The diffusion constant D was found using the expression $D = ½v_F^2(\tau_i^{-1}+\tau_*^{-1})^{-1}$, which considers point disorder for intervalley scattering and charged-impurity disorder for intravalley scattering.[37] From the analysis we obtained values of $4.4\times10^{-12}$ s, $9.6\times10^{-12}$ s and $1.4\times10^{-14}$ s for $\tau_\phi$, $\tau_i$, and $\tau_*$, respectively, which correspond to lengths $L_\varphi \approx 540$ nm, $L_i \approx 800$ nm, and $L_* \approx 10$ nm. Of course these values are only approximate since they strongly depend on carrier density. Magneto-conductivity was also strongly affected by universal conductance fluctuations. In any case, our results can be compared with those from weak localization investigations of graphene on $SiO_2$.[33,34] In both cases intravalley scattering times were approximately two orders of magnitude smaller than dephasing and intervalley scattering times. Based on [35] we conclude that our system was in the electron localization regime. Calculated ratios of $\tau_\phi/\tau_* > 10$ confirm such an electron localization. However, the ratio $\tau_\phi/\tau_i \approx 0.5$ is close to the theoretical value separating localization and anti-localization regimes. We suppose that this was due to the relatively low carrier density below $1\times10^{12}$ cm$^{-2}$ used in our experiment.

## VI. Conclusion

In conclusion we have presented a study of the magneto-transport of exfoliated graphene prepared on a specially grown GaAs/AlAs substrate. It is possible to control the carrier density in graphene by the back-gating action of the doped GaAs substrate despite a much lower insulating barrier than that provided by $SiO_2$ on silicon substrates. Although the accessible voltage range is limited by the onset of gate leakage, the characteristic features of graphene can be observed, since the higher dielectric constant of GaAs/AlAs makes the tuning by a back-gate more efficient, and since - al-ready in the as-prepared state - the charge neutrality points lie within the accessible gate voltage range. Apart from these differences graphene on GaAs behaves very similarly to graphene on $SiO_2$, as the quantum Hall voltage features at high magnetic field and the weak localization effects at low field have shown. The observed deviation of the Hall quantization from the exact $h/2e^2$ value at fields of 6 Tesla and below is probably caused by the coexistence of electron and hole puddles occurring at low carrier densities in the vicinity of the charge neutrality point. Since also in graphene prepared on $SiO_2$, no Hall quantization as precise as that observed for epitaxial graphene has been reported to date, our findings underpin the fact that it will be a non-trivial task to prepare exfoliated graphene in such a way that it can serve as a precise resistance reference at very low magnetic fields. There it would allow its unique advantages over an established quantum Hall system like a GaAs based two dimensional electron system to be fully exploited. However, applications of graphene which is tailored to the substrate not aiming at such ultra-precision measurements will hopefully benefit from the findings reported in this paper.

**Acknowledgment**

This research has received funding from the European Community's Seventh Framework Programme, ERA-NET Plus, under Grant Agreement No. 217257. We further would like to thank T. Dziomba for carrying out AFM measurements, R. Stosch, and S. Wundrack for Raman spectroscopy investigations, P. Hinze for SEM measurements, and G.-D. Willenberg, J. Schurr, and D. Weber for their fruitful discussions.

[27] S. Adam, E.H. Hwang, V.M. Galitski, and S. Das Sarma, "A self-consistent theory for graphene transport," *Proceedings of the National Academy of Sciences*, vol. 104, 2007, pp. 18392-18397.

[28] E.H. Hwang, S. Adam, and S.D. Sarma, "Carrier Transport in Two-Dimensional Graphene Layers," *Physical Review Letters*, vol. 98, 2007, p. 186806.

[29] V. Galitski, S. Adam, S. Das Sarma, Statistics of random voltage fluctuations and the low-density residual conductivity of graphene, Physical Review B. 76 (2007) 245405

[30] Y. Zhang, V.W. Brar, C. Girit, A. Zettl, and M.F. Crommie, "Origin of spatial charge inhomogeneity in graphene," *Nature Physics*, vol. 5, Aug. 2009, pp. 722-726.

[31] A. Deshpande, W. Bao, F. Miao, C. Lau, and B. LeRoy, "Spatially resolved spectroscopy of monolayer graphene on SiO2," *Physical Review B*, vol. 79, May. 2009, p. 205411.

[32] S. Branchaud, A. Kam, P. Zawadzki, F.M. Peeters, and A.S. Sachrajda, "Transport detection of quantum Hall fluctuations in graphene," *Physical Review B*, vol. 81, Mar. 2010, p. 12406(R).

[33] F.V. Tikhonenko, D.W. Horsell, R.V. Gorbachev, and A.K. Savchenko, "Weak Localization in Graphene Flakes," *Phys. Rev. Lett.*, vol. 100, Feb. 2008, p. 56802.

[34] R.V. Gorbachev, F.V. Tikhonenko, A.S. Mayorov, D.W. Horsell, and A.K. Savchenko, "Weak Localization in Bilayer Graphene," *Phys. Rev. Lett.*, vol. 98, Apr. 2007, p. 176805.

[35] F.V. Tikhonenko, A.A. Kozikov, A.K. Savchenko, and R.V. Gorbachev, "Transition between Electron Localization and Antilocalization in Graphene," *Physical Review Letters*, vol. 103, Nov. 2009, p. 226801.

[36] J. Eroms and D. Weiss, "Weak localization and transport gap in graphene antidot lattices," *New Journal of Physics*, vol. 11, 2009, p. 095021.

[37] J. Moser, H. Tao, S. Roche, F. Alzina, C.M. Sotomayor Torres, A. Bachtold, "Magnetotransport in disordered graphene exposed to ozone: From weak to strong localization", *Physical Review B*, 81 (2010) 205445.